# Resilience Evaluation Framework for Integrated Critical Infrastructure-Community Systems under Seismic Hazard

Li Sun[1,*]; Bozidar Stojadinovic[2]; Giovanni Sansavini[3]

[1]Chair of Structural Dynamics and Earthquake Engineering, Institute of Structural Engineering, ETH Zurich, Switzerland, sun@ibk.baug.ethz.ch

[2]Chair of Structural Dynamics and Earthquake Engineering, Institute of Structural Engineering, ETH Zurich, Switzerland, stojadinovic@ibk.baug.ethz.ch

[3]Laboratory of Reliability and Risk Engineering, Institute of Energy Technology, ETH Zurich, Switzerland, sansavig@ethz.ch

ABSTRACT: Seismic resilience of civil infrastructure systems is an essential property of modern communities. In this paper, an agent-based modeling framework to evaluate the seismic resilience of an integrated system comprising the community and its civil infrastructure systems is proposed. Specifically, an agent-based model of the recovery process of civil infrastructure systems is incorporated into a previously developed compositional supply/demand seismic resilience quantification framework. The proposed model represents the behavior of the operators of civil infrastructure systems as they strive to recover their functionality in the aftermath of an earthquake as well as their mutual interactions and the interactions with the community they provide services to. A case study of the seismic resilience of a virtual system comprising the electric power supply system, the transportation system and the community (EPSS-TS-Community system) is conducted using the proposed framework. A parametric investigation is carried out to examine the effect of different earthquake magnitude scenarios as well as different behaviors of the involved agents and their interaction on the seismic resilience of the EPSS-TS-Community system. It was demonstrated that the proposed agent-based modeling approach is effective in representing the interactions among different participants in the recovery process. It was also revealed that timely and well-planned intervention in the recovery process can be very effective in alleviating the post-earthquake lack of resilience resulting from the insufficient supply of civil infrastructure service to meet the community demands. Therefore, the proposed framework could be employed to formulate the recovery trajectory of the intertwined socio-technical system subjected to different earthquake scenarios. The interplay among different agents, as well as the interdependence between the civil infrastructure systems is found to profoundly shape the recovery path for this integrated EPSS-TS-Community system.

*Corresponding author
*Email address:* sun@ibk.baug.ethz.ch (Li SUN),
Preprint, on *ASCE Journal of Infrastructure Systems*

## Introduction

The Civil Infrastructure Systems (CISs) function as the arteries of modern urban communities (Mieler et al. 2015). These infrastructure systems are becoming more integrated and interdependent through sharing of resources and exchange of information (Kröger and Zio 2011, Helbing 2013). Such interdependence enables more efficient functioning of the CISs, but simultaneously renders the intertwined CISs less redundant and more vulnerable to disruptive events such as natural hazards, terrorist attacks, and random technical errors (Albert et al. 2004, Zhao et al. 2010, Kawashima 2012). Local damage can be exacerbated by cascading through one or more CISs and lead to a severe global failure of CISs (Dueñas-Osorio and Vemuru 2009, Buldyrev et al. 2010, Zio and Sansavini 2011). According to the field observations from recent catastrophic earthquake disasters around the world, CISs usually struggle to recover from the seismic damage they suffered (Hollnagel and Fujita 2013, Hwang et al. 2015). Moreover, the stagnant recovery of one CIS often hinders the restoration of many other CISs, as well as the rescue and evacuation of people in the affected areas (Kawashima et al. 2009, Lekkas et al. 2012, Krishnamurthy et al. 2016). Recovery and restoration are, indeed, critical post-disaster phases, particularly if the aims are to achieve a higher level of system performance as compared to the pre-disaster state (Fang and Sansavini 2017).

Many studies on the behavior of CISs in urban communities under stress caused by natural hazards in general and earthquakes in particular have been carried out over the past decades. The engineering models of CIS and urban community resilience (Hosseini et al. 2006) are usually based on the post-event functionality deterioration and recovery model proposed by Bruneau and colleagues at MCEER (Bruneau et al. 2003). The state-of-the-art models attempted to shape the restoration trajectory of CISs with the set of parameterized recovery functions, on either system or component levels (Decò et al. 2013, Sun et al. 2015). Though practical, this paradigm is struggling to account for the case-specific socio-technical characters of each individual Civil Infrastructure-Community System (CICS), and more importantly, the dynamic interactions among the restoration of service from different CISs throughout the intertwined recovery campaign. A smaller number of studies address the seismic resilience of interdependent CISs. They show that increasing robustness and reparability, two basic elements of resilience, improves the seismic behavior of coupled CISs (Farr et al. 2014). Considerable research focuses on CIS robustness. Bashan et al. (2013) pointed out that the failure of a tiny fraction of the nodes of any sub-system can potentially fragment the globally integrated CIS. Against that backdrop, Brummitt et al. (2012) indicated that the interdependence among CISs should be set on an optimal level in order to avoid a devastating cascade among interconnected CISs. Correspondingly, the research conducted by Schneider et al. (2013) also revealed that the nodes with high betweenness play a significant role in cascades within coupled networks.

In this paper, a framework to model the post-earthquake recovery of an integrated CICS is developed by combining an agent-based component and system recovery model (Sun 2017) with the compositional (bottom-up, component to system) supply/demand framework for assessing the seismic resilience of a system comprising the community and a single CIS by quantifying functionality loss in the absorption stage and tracking the functionality restoration trajectory throughout the recovery stage (Sun et al. 2015, Didier et al. 2015). Specifically, the Electric Power Supply System (EPSS), the Transportation System (TS) and the urban community they serve, are examined. The agent-based model (ABM) of the post-earthquake recovery process is extended to capture the post-earthquake recovery path of the community served by interdependent EPSS and TS. Three individual agents, the Operators of the EPSS and the TS, and the community Administrator are defined by attributes that model their behaviors during the post-earthquake recovery process. More importantly, the rules for the interaction among the agents are also specified. The proposed framework is exemplified using a virtual EPSS-TS-Community system.

The recovery trajectory of the system is computed for different earthquake scenarios. The influence of the interdependence between EPSS and TS on the systemic resilience of the community they serve is highlighted. The impact of different agent behavior characteristics on the systemic resilience is also investigated and discussed.

## Agent-Based Seismic Resilience Modeling Framework

Modern CICS is a heterogeneous and intertwined socio-technical network. Its natural hazard resilience is a dynamic process that depends on many factors (Cimellaro et al. 2016). In a typical seismic event, the components of both EPSS and TS and the elements of the built environment of the urban community will be damaged to some degree. Note that unlike biological or financial systems, the physical damage of CISs usually cannot recover spontaneously (Majdandzic et al. 2013). Instead, their functionality can only be restored if apt external recovery measures are executed. The goal of the proposed agent-based modeling framework is to represent these measures and the interactions among them during the pose-earthquake recovery process, and quantify the resulting seismic resilience of the integrated CICS.

### Agent-Based post-earthquake recovery model of the CICS

An integrated CICS is, in this model, represented by the EPSS, the TS and the community built environment. Seismic resilience of the EPSS-TS-Community system is modeled using the compositional supply/demand resilience quantification framework (Didier et al. 2018a). The state of the electric power supply and demand in the EPSS-TS-Community system is modeled through the earthquake damage absorption and recovery phases. The absorption phase is a relatively short period immediately following an earthquake when the EPSS-TS-Community systems is accumulating damage (direct and cascading) and finding a new equilibrium at a substantially lower functionality level. Damage to the EPSS and TS components (substations and bridges, respectively) and the Community built environment components (residential buildings where the inhabitants are, clustered in the residential sectors of the Community, and offices, factories and other structures that support the economy of the region, clustered in the industrial sectors of the Community) is computed using seismic vulnerability functions that express the probability that each component will retain a certain portion of its functionality conditioned on the intensity of the earthquake ground motion at its location. In this model it was assumed that the power transmission lines and the roads between bridges do not incur earthquake damage and thus suffer loss of functionality. The decrease of power supplied by the EPSS and the decrease of the power consumed by the community was assessed in proportion to the loss of component functionality using a model of EPSS operation specific for a seismic contingency.

After the absorption phase, the EPSS-TS-Community system enters a considerably longer recovery phase. This phase involves a lengthy repair effort to reestablish the full functionality of the damaged EPSS and TS components and a parallel effort to repair the damaged buildings, to restore community functions and to ensure the distribution of the essential services, such as electric power. In this model, the restoration of the components of the community built environment is represented using recovery functions (RFs) (e.g. HAZUS, 2015), which are cumulative distributions of the probability that a building is repaired after a certain period of time conditioned on its damage state. However, in contrast to a model used in (Sun et al. 2015, Didier et al. 2015), the recovery process of the components of the EPSS and TS is modeled using an agent-based paradigm to plastically represent the actual repair procedure undertaken within each CIS and the interdependencies among the recovery sequences. The underlying assumption is that the post-earthquake repair of the components of the community built environment is independent of the recovery of CISs. Agent-based models could also be used to simulate the recovery of the community built environment and include the interdependence between this effort and that or the CIS recovery: this was not done in this model to make the simulations manageable and to emphasize the CIS recovery interdependencies.

Agent-based model (ABM) is a well-known strategy (e.g. O'Sullivan & Haklay 2000) employed to capture the dynamic behavior of a broad range of social, economic and ecological systems where the interaction of constituents (autonomous agents) depends highly on the particular state of the system and the interaction among the constituents. It can, therefore, deliver a nuanced model of the post-earthquake recovery of interacting CISs and the community built environment, accounting for different strategies and capabilities of all players involved in the recovery process.

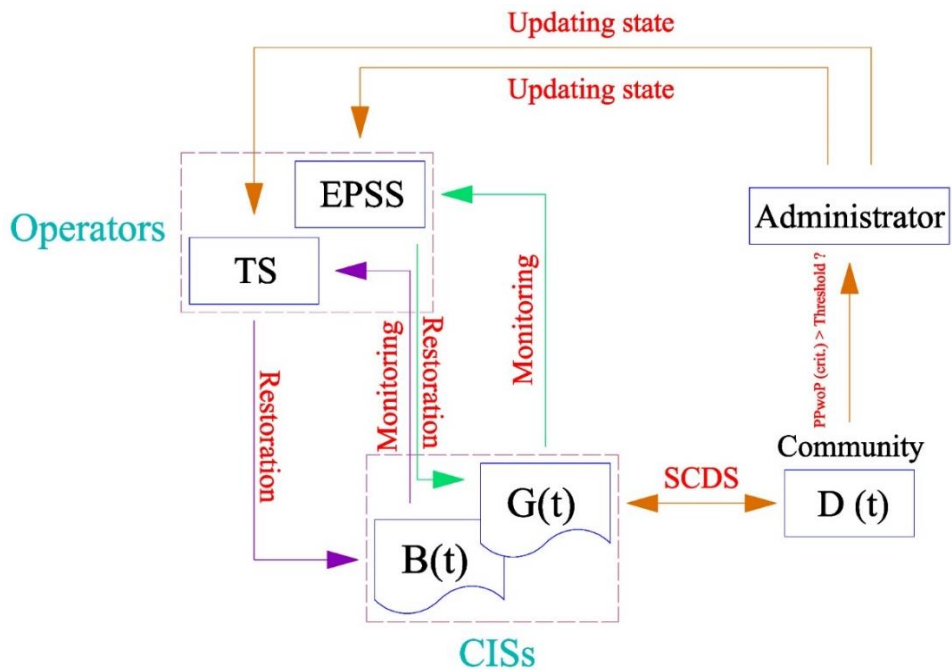

Fig. 1. Illustration of the proposed agent-bases seismic resilience modeling framework.

Accordingly, three principal players, the **EPSS Operator**, the **TS Operator** and the community **Administrator** are defined as separate agents, as illustrated in Fig. 1. Immediately after an earthquake, the community and the CIS operator agents enter a phase of damage absorption. As some information about the damaged state of the CISs is obtained, the **EPSS Operator** and **TS**

**Operator** agents mobilize their resources and start the recovery process of their CISs. In this model, CIS operators act independently, following their own repair plans based on their own priorities and resource constraints. However, due to the interdependence among different CISs, the recovery of service provided by a single CIS affects, and is, in turn, influenced by the restoration of the other CISs, as well as by the recovery of the community. Crucially, implementation of the EPSS recovery plan depends on the availability of the TS: EPSS repair crews can only reach the substations by driving along the instantaneously available roads and bridges, which are simultaneously undergoing repairs. This is the essential interaction captured by the proposed ABM. Therefore, in the extension of ABM for the single CIS (Sun 2017), the **EPSS Operator** repair crew computes the shortest path along the available TS links between its current location and the location of the next substation on the repair schedule. Initially, the model assumes that the post-disaster recovery priorities of EPSS and TS are governed by their own business needs and repair costs and effort, not necessarily by the recovery requirements of the community they serve. Therefore, the recovery process may need to be overseen and governed externally to better serve the community needs. Typically, an emergency management coordination entity assumes such a role (Liel et al. 2013). The **Administrator** agent is, therefore, defined in this model to coordinate the CIS recovery campaign and optimize the outcome with respect to predefined community priories.

The behavior of the three agents is defined using random variables called attributes, deterministic parameters, and sets of pre-defined repair crew deployment prioritization plans, as follows:

> **EPSS Operator** is an agent whose behavior is described by three attributes: $V_e$, $E_{eg}$, and $E_{ed}$. Specifically, $V_e$ describes the average travelling speed of the EPSS repair crews between repair locations, while $E_{eg}$ and $E_{ed}$ quantify the repair efficiency (as the percentage of functionality restoration per day) for the generation substations and the distribution substations of EPSS, respectively. A parameter that defines the duration of agent resource mobilization from the time of the event to repair crew deployment, $t_{0e}$, is also specified. The generation substations are ordered by their damage grade, from least to most seriously damaged ones, thus forming a repair schedule. Such repair sequence enables quick recovery of the least damaged portions of the EPSS system first, provides more options for power dispatch management early on, and maximizes the amount of supplied power, thus potentially maximizing the profit of the EPSS (Sun 2017). Correspondingly, the distribution substations are ranked according to the population of the communities they serve, from largest to the smallest. Once a repair crew is done with restoring functionality to an EPSS component, it travels to the next-ranked component in the current repair schedule.

**TS Operators** is an agent whose behavior is defined by two attributes $V_b$, and $E_b$. Similar to EPSS Operator, $V_b$ refers to the travelling speed, while $E_b$ quantifies the bridge restoration efficiency of the repair crew. A resource mobilization time $t_{0b}$ is also specified. The damaged bridges in the TS are ranked using their betweenness centrality value (Brandes 2001). Betweenness centrality quantifies the number of times a bridge acts as one link along the shortest path between any other two nodes inside the TS treated as a graph (Freeman 1977). The damaged bridges are ranked by their betweenness centrality in descending order to form a repair schedule. Such repair priority is adopted on the grounds that a bridge with higher betweenness centrality is more prone to become a traffic flow bottleneck within the TS. In addition, it is also assumed that the TS repair does not depend on the electric power provided by the EPSS. Once a repair crew is done with restoring a bridge to its full functionality, it travels to the next-ranked bridge on the repair schedule.

**Administrator** is an agent that monitors the community recovery process using a set of CICS resilience measures. In this model, the percentage of people without power (*PPwoP*) (Sun et al. 2015, Sun 2017) is chosen as the resilience measure. *PPwoP(t)* is the ratio of the number of people whose demand for electric power is not met to the total number of people in the EPSS-TS-Community system at each instant *t* of the recovery process simulation. This measure directly addresses the affected population in the residential sectors of the Community, but does not measure the losses caused by the inability to supply electric power to the industrial sectors of the Community. The community recovery performance objectives are formulated in terms of exceeding threshold values of the selected resilience measures at certain intervals, following (SPUR 2009). In this model, the Administrator agent determines if the rate of community recovery is satisfactory or not by comparing the *PPwoP* to a threshold value at one point in time after the earthquake called the Resilience Check Time (RCT). If the rate of recovery is not satisfactory at this time, i.e. if the attained *PPwoP(RCT)* is larger than the pre-set threshold, the **Administrator** agent state is switched from "inactive" to "activate". It intervenes by forcing the **EPSS Operator** and the **TS Operator** agents to perform a one-time irreversible adjustment of their attributes and repair schedules. Namely, the travelling speed and the repair efficiency attributes of both the **EPSS Operator** and the **TS Operator** agents are incremented to speed up the recovery process, and the repair schedules are updated to prioritize the recovery of the community (e.g. by repairing the most heavily damaged generation substations first) at increasing costs to the CIS operators.

### *Monte Carlo simulation of post-earthquake recovery of the CICS*

A simulation starts by defining the CICS. This includes the specification of the location, occupancy and the seismic vulnerability and recovery functions for the community buildings and the specification of the locations and vulnerability of the EPSS and TS components, the location of the EPSS and TS repair centers, agent resource mobilization times and probability distributions of the agent attributes, as well as the description of the EPSS and TS networks, their topology and operation, and setting the CICS post-earthquake recovery performance goals (Sun 2017). An earthquake scenario is defined by the earthquake magnitude and the location of the earthquake hypocenter. The damage states of the components of the EPSS, the TS and the community are evaluated using the corresponding seismic vulnerability functions, contingent on the ground motion intensity measures computed using ground motion prediction equations. This data is used to determine the EPSS and TS repair schedules.

In a Monte Carlo simulation framework, a single simulation instance is set by randomly generating the values of the five agent attributes from the specified distributions of the agent attribute values. The functionality of any damaged generation substation $n$ (out of $N_g$ damaged generation substations) before and after the earthquake is denoted as $f_{o,n}$ and $f_{d,n}$, respectively, and measured in terms of the generated electric power. Thus, the damage severity

$$S_n = 100 \times (1 - f_{d,n}/f_{o,n}) \ (\%) \tag{1}$$

of this substation is its functionality loss normalized by its original functionality level. Thus, $S_n$=100% means a substation is out of service, $S_n$=0% means that a substation is fully operational, and a value in between signifies that a distribution substation is operating but partially in proportion to the number of generators connected to this substation. The repair time for this substation is computed as $S_n/E_{eg}$, a ratio of its damage severity and the repair rate of the generation substation repair crew, and is measured in days. The travelling time between the damaged generation substations $n$-1 and $n$ (for $n \in [2, N_g]$, with $n$=0 denoting the location of the repair center) is calculated as $SD_n/V_e$, where $SD_n$ is the shortest distance between substations $n$-1 and $n$, determined according to the instantaneously available TS, and $V_e$ is the travelling speed parameter of the EPSS repair crew. Hence, for any damaged substation $n$ on the repair schedule, the time of repair crew arrival $t_n$ and the time of repair completion $r_n$ is iteratively computed as:

$$t_1 = t_{0e} + SD_0/V_e; \quad for \; n = [2, N_d] \quad \begin{cases} r_n = t_n + S_n/E_{eg} \\ t_n = r_{n-1} + SD_n/V_e \end{cases} \quad (2)$$

where $t_{0e}$ is the EPSS resource mobilization time and $SD_0$ is the shortest distance between the repair center and the first generation substation to be repaired. The seismic restoration campaign evolves though the discrete time steps $t_n$ and $r_n$ as the repair of the generation substations proceeds along the priority list until all of them are repaired. The generation capacity $G(t)$ [MW] of the EPSS-Community system at time $t$ is a sum of the available power generation functionality of the generation substations (Sun 2017). This model assumes that the generation substations of the EPSS can receive electric power from the rest of the interconnected grid without restrictions, and, therefore, that power markets do not constraint the resources to the restoration process. The recovery of the $N_d$ damaged distribution substations is modeled in the same way, the only difference being that partial repairs, possible for generation substations, are not possible for distribution substation. The effort to repair the community buildings is simulated in parallel using recovery functions to compute the state of building at each point in the simulation and, using building occupancy data, estimate its potential for power consumption. The demand for electric power posed by the Community at time $t$, $D(t)$ [MW], is computed by summing up the consumption capability of the community buildings in the attained state of damage and repair, considering they are independent consumers of electricity. Together, the available functionality of the eclectic power generation and distribution substations and the electric power demand posed by the community determines the ability of the EPSS to deliver power to the community at time $t$, $DP(t)$ [MW]. The dispatch of the generated electric power is computed using a seismic contingency dispatch strategy (SCDS) designed to balance the EPSS supply and demand, protect the functioning components of the EPSS, and supply functioning consumers in some order of priority (Sun 2017). A condition when the community demand for electric power exceeds the power delivered by the EPSS at each distribution substation represents a power deficit and indicates the EPSS-TC-Community system lacks resilience. An instantaneous measure of such Lack of Resilience is *PPwoP(t)*.

Recovery of the TS is modeled using the same approach. Similar to distribution substations, a damaged bridge is available again to the traffic only if it is fully repaired. The repair priority of bridge $b$ of $N_b$ damaged bridges is assessed by computing its betweenness as:

$$\beta(b) = \sum_{b \neq s \neq t \in B} \frac{\sigma_{st(b)}}{\sigma_{st}} \tag{3}$$

where the set of vertices **B** contains all bridges in the TS, $\sigma_{st}$ is the total number of shortest paths from node $s$ to node $t$, and $\sigma_{st}$ *(b)* is the number of those paths that traverse bridge $b$. A total of fully functional bridges *B(t)* at time $t$ is computed in a simulation to measure the recovery of the TS. Each MC simulation ends when all components of the EPSS-TS-Community system are restored to their pre-earthquake functionality.

## Case Study

The EPSS-TS-Community system used in this case study is shown in Figure 2. The EPSS (Fig. 2(a)) is a portion of the IEEE 118-node Benchmark System (Christie 1993). It comprises 15 electric power generation substations (red squares) and 19 electric power distribution substations (blue circles). The role of the generation substations is to connect the electric power generators to the high-voltage power grid, while the role of the distribution substations is to connect low voltage electric power to consumers. The population centers are located near the distribution substations. Detailed data on the substation power capacities and the population they serve are provided in (Sun 2017). The TS (Fig. 2(b)) is then designed to connect the population centers with some redundancy.

A total of 32 bridges (red dots) are distributed along the TS links that represent the roads. The EPSS repair center is located near node 42 while the TS repair center is located near node 35. The vulnerability functions for the EPSS substations, the TS bridges and the community buildings, as well as the community building recovery functions, are specified in (Sun 2017). They are consistent with the typical inventory of modern infrastructure systems and communities that is designed for low probability of collapse and loss of life for design-level seismic hazard, but is likely to incur significant damage in such events. Of five electric power seismic contingency dispatch strategies investigated in (Sun 2017), SCDS 1, which prioritizes the supplying electric power to communities that have the largest demand, is used in this case study.

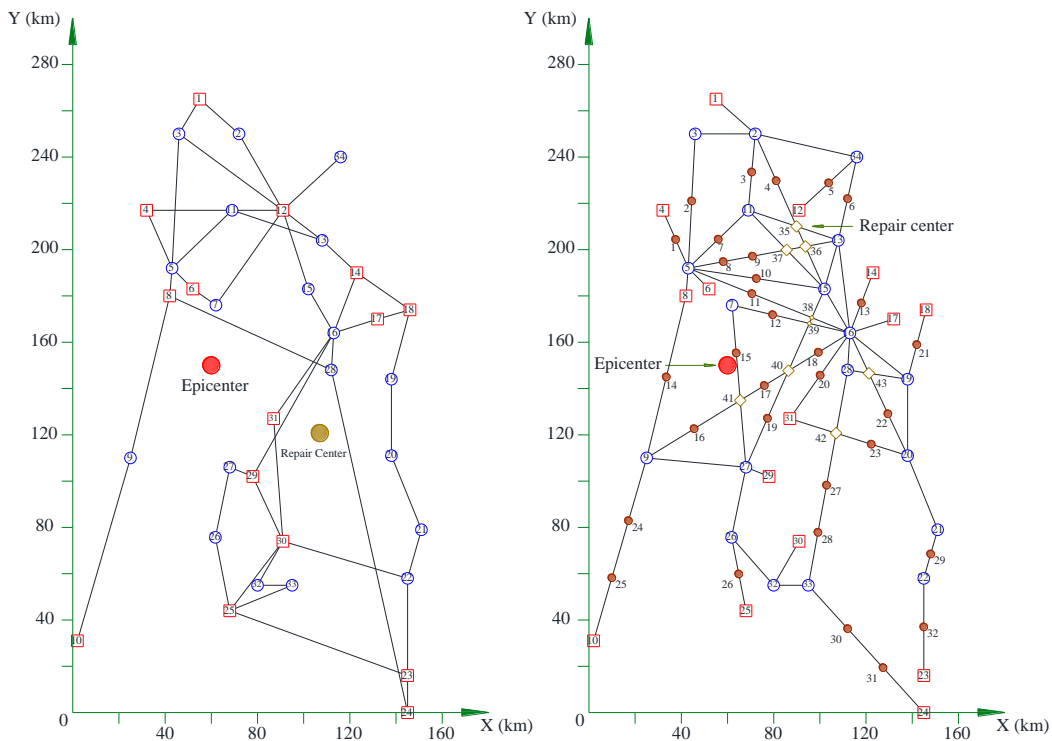

Fig. 2. The EPSS-TS-Community system investigated in the case study (Fig. 2(a) EPSS, Fig. 2(b) TS).

The seismic hazard is modeled by locating the earthquake hypocenter close to the geographic center of the EPSS system, as shown in Figure 2. This hypocenter location is not changed in this case study to control the computational effort. The magnitude of the earthquake ($M$) is associated to the occurrence probability using the bounded Gutenberg-Richter law (the seismic hazard curve) with parameters *a*=4.4 and *b*=1 (Kramer 1996), indicating a design hazard (475 return period) earthquake magnitude is approximately 7. The intensity of shaking at each EPSS or community component site, measured using peak ground motion displacement, velocity and acceleration values is computed using the ground motion attenuation relations proposed by Campbell & Bozorgnia (2008). The geographic size of the EPSS-TS-Community system is scaled down by a factor of 5 to cover, roughly, and area of 32x50km. Thus, significant ground shaking due to earthquakes with magnitudes between 4 and 7 is expected at most of the EPSS, TS and community component locations. Clearly, such increase in the EPSS-TS-Community density will result in increased damage and prolonged recovery time, but will also exacerbate the dependency of the EPSS recovery on the TS bridge repairs, helping to demonstrate the features of the proposed ABM.

Table 1. Attributes and parameters of the **EPSS Operator**, **TS Operator** and **Administrator** agents.

| Attribute | Distribution Type | Lower Limit | Upper Limit |
|---|---|---|---|
| **EPSS Operator** | | | |
| $V_e$ (km/h) | Uniform | 8 | 10 |
| $E_{eg}$ (1/day) | Uniform | 20% | 40% |
| $E_{eg}$ (1/day) | Uniform | 50% | 100% |
| $t_{0e}$ (day) | Constant | 1.5 | |
| **TS Operator** | | | |
| $V_b$ (km/h) | Uniform | 7 | 8 |
| $E_{eb}$ (1/day) | Uniform | 40% | 80% |
| $t_{0b}$ (day) | Constant | 2.0 | |
| **Administrator** | | | |
| $PPwoP$ (%) | Constant | {100, 10, 20, 30} | |
| RCT (days) | Constant | 3 | |

The parameters of the **EPSS Operator**, **TS Operator** and **Administrator** agents are shown in Table 1. For instance, the $V_e$ attribute of the EPSS Operator follows a uniform probability distribution and takes values between 8 and 10 km/h, consistent with the length scaling of the EPSS network. This case study was conducted assuming that the **EPSS Operator** has two repair crews, one for the generation and the other for the distribution substations (their repair efficiencies are thus different), and that the **TS Operator** has a single repair crew. This assumption further emphasizes the effect of the interaction between the agents during the recovery process in this case study, but also elongates this process. The distinction between repairs for lightly and moderately damaged bridges and rebuilding for severely damaged bridges is modeled by changing the mode of operation of the TS repair crew. Namely, the duration of the repairs is computed using the TS Operator repair efficiency parameter, while the duration of the much longer rebuilding is specified by a random variable with a lognormal distribution whose parameters are specified in Table 2.

Table 2. Parameters of the lognormal distribution of the rebuilding time for severely damaged bridges.

| Administrator | Mean (Days) | Std. (Days) |
|---|---|---|
| inactive | 150 | 90 |
| active | 90 | 60 |

Four values of the recovery performance *PPwoP* threshold of the **Administrator** agent shown in Table 1, were investigated in this case study, while the resilience check time (RCT) is set to be 3 days after the earthquake. If the rate of the EPSS-TS-Community recovery is too slow, the **Administrator** agent becomes active and forces a change of the **EPSS Operator** agent repair schedule by reordering the remaining damaged generation substations by descending damage severity. In this case study, the repair schedule for the distribution substations and the bridges remained the same. In addition, the repair efficiency of the EPSS Operator repair crews is doubled (but not to exceed 100%), and the traveling speed of the repair crews is increased by 10%. Similarly, the traveling speed of the TS Operator repair crew is increased by 10%, its repair efficiency for lightly and moderately damaged bridges is doubled and the rebuilding time for severely damaged bridges is shortened by changing the lognormal probability distribution parameters as shown in Table 2.

Two cases are investigated separately. First, the Monte Carlo simulations were conducted without the **Administrator** agent, in order to observe the behavior of the system where Community performance objectives are neglected and the only interaction is between the **EPSS Operator** and the **TS Operator** agents that impacts the travel time of the EPSS repair crews. Second, the simulations were conducted with the **Administrator** agent, allowing the Community to interact with the EPSS and TS Operators and assert its post-earthquake recovery performance objectives. The effects of earthquake magnitude and agent interaction and were investigated in each case, and compared to the baseline created by assuming the TS is not damaged, thus idling the **TS Operator** agent and eliminating agent interactions.

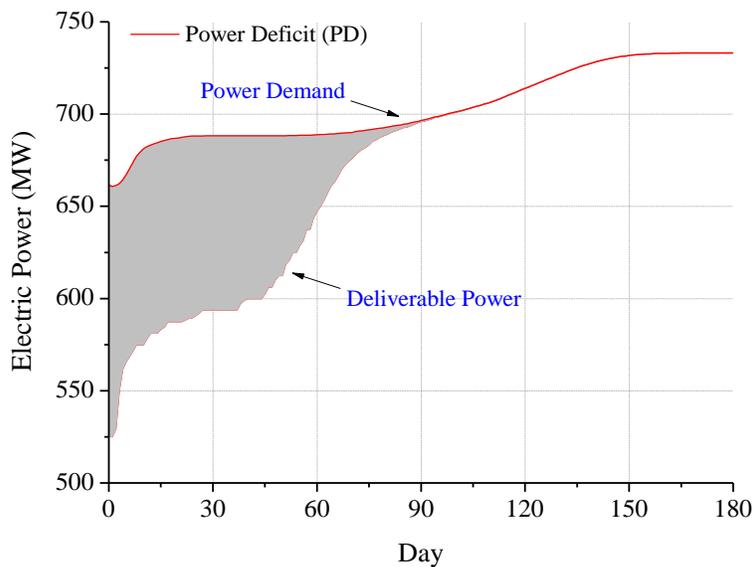

Fig. 3. Median of deliverable power (*DP*) and power demand (*D*) in the EPSS-TS-Community system for an interacting M=7.5 earthquake scenario case study simulations without the **Administrator** agent.

In each Monte Carlo simulation, the statistics of the resilience indicators are derived from 2,000 instances. The maximum duration of the post-earthquake recovery process was set to 360 days, but most of the simulation instances were complete much sooner. For example, graphs of the median values of the EPSS deliverable power $DP(t)$ and the community power demand $D(t)$ in the M=7.5 earthquake scenario considering **EPSS Operator** and **TS Operator** agent interaction without the **Administrator** agent are shown in Figure 3. The shaded area represents the power deficit and signifies the lack of resilience of the EPSS-TS-Community system. The $PPwoP(t)$ resilience measure is an instantaneous value of the power deficit, expressed in terms of the number of people who could use electric power but are not receiving it. Note that the graphs start at the time of the earthquake event, assuming the damage absorption phase is instantaneous. Immediately after the earthquake, the median community demand decreased to about 660 MW as the earthquake damage is absorbed in the community. Meanwhile, the median power generation capacity drops to 575 MW. Further, due to failure of distribution substations, the median deliverable power decreased to 531 MW. Thus, the EPSS is not able to satisfy the demand, resulting in a lack of the resilience of the EPSS-TS-Community system. The level of deliverable power remains almost unchanged over the first three days and then starts to increase. The gap between the $DP(t)$ and $D(t)$ reduces, and disappears on the 102nd day after the earthquake. However, it took 165 days to restore the community power demand to the pre-disaster level of 733 MW. The electric power supply capacity of the EPSS was able to follow this increase in demand without problems.

*Case without the Administrator agent*

The effect of interaction of the EPSS Operator and the TS Operator agents is evident in Fig. 4, where the recovery of EPSS power generation capacity $G(t)$ to its 900 MW pre-earthquake level is shown for earthquake magnitude of 6, 6.5, 7 and 7.5 scenarios in the baseline (no TS damage) and interacting (TS is damaged) simulations. The recovery process in the interacting scenarios is much (about 4 times) longer than in the baseline scenario because the EPSS repair crews are idle while they wait for the TS repair crew to complete various bridge repairs. In particular, a bottleneck in the TS system created by long repair and rebuilding of a few critical bridges was alleviated only after about 50 days in magnitude 6.5, 7 and 7.5 scenarios. On the other hand, bridge damage in the magnitude 6 scenario was neither severe nor wide-spread, enabling a higher rate of EPSS power generation recovery. This result indicates that post-earthquake recovery simulations without considering the interaction among the community CISs may produce over-optimistic estimates of the recovery times, more so for more intense earthquakes.

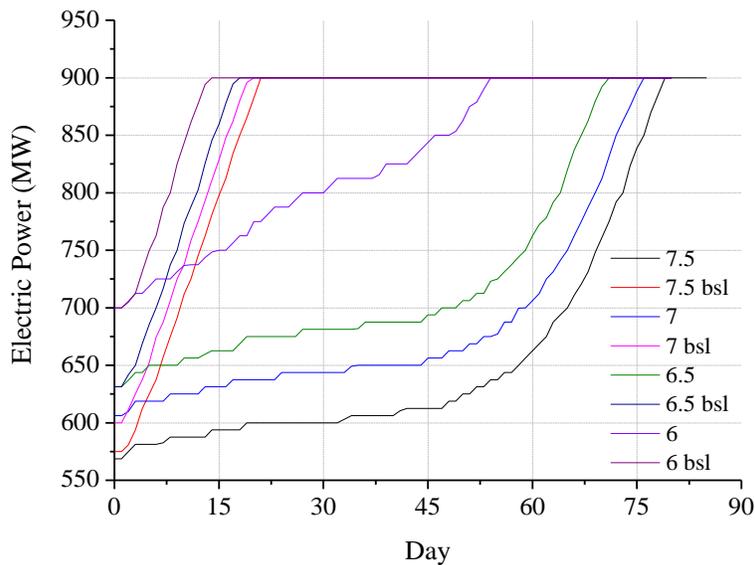

Fig. 4. Median generation capacity (*G*) recovery curves in the baseline and interacting case study scenario simulations without the **Administrator** agent.

The effect of randomness in MC simulations of the EPSS-TS-Community post-earthquake recovery is illustrated in Fig. 5, where probability density histograms of the generated power values computed in the 2000 simulation instances of the interacting M=7.5 scenario are shown together with the normal distribution graphs derived using the mean and the standard deviation of the corresponding datasets. The data collected on the first day after the earthquake is uniformly distributed. The data collected on the 60th day after the earthquake is truncated because the EPSS generation capacity recovers fully by that time in a significant number of simulations, resulting in the median of the data being larger than the mean. This test can be used to identify the occurrence of a significant number of early recovery completions and indicate the need to further analyze the dataset.

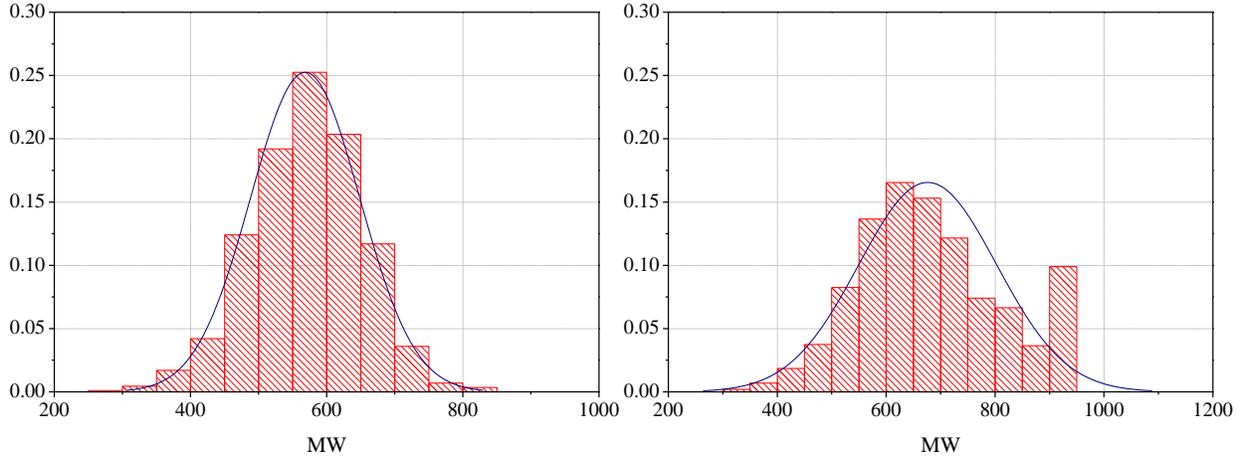

Fig. 5. Histogram of EPSS generation capacity (*G*) in the interacting M=7.5 scenario case study simulations without the **Administrator** agent (Fig. 5(a) First day, Fig. 5(b) 60<sup>th</sup> day).

The evolution of the median percentage of fully functional bridges *B(t)* in the interacting scenarios with four different magnitudes is shown in Fig. 6. The number of operational bridges immediately after the earthquake was 21.88%, 25%, 28.13% and 37.5% for magnitude 7.5, 7, 6.5 and 6 scenarios, indicating that the damage to the TS was widespread. The recovery process starts from the bridge with the highest betweenness centrality and moves to the next-nearest bridge, showing a rate consistent with the repair efficiency of the TS repair crew. However, a significant slowdown in the TS recovery process is caused by a few severely damaged bridges that require a long time to rebuild (Table 2). The location of these bridges with respect to the earthquake hypocenter and the structure of the TS network are the reasons for fairly similar mean duration of full TS recovery (78 days for M=7.5 and 76 days for M=6 scenarios). Simulations with the location of the earthquake hypocenter also varied (e.g. randomly select along hypothetical fault lines) would reveal the effect of TS network structure on its recovery time.

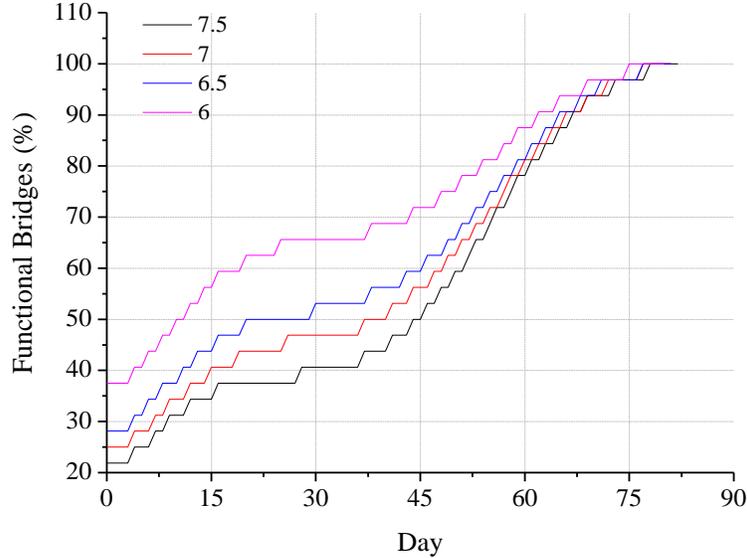

Fig. 6. Median of the fraction of functioning bridges in the TS in four interacting earthquake case study scenarios without the **Administrator** agent.

Fig. 7 illustrates the evolution of the median values of the *PPwoP* resilience measure of the EPSS-TS-Community system. A comparison of the baseline and interaction scenarios in Fig. 7(a) shows that the TS plays a critical role in recovery of the EPSS-TS-Community system: in more intense earthquake scenarios (M=7.5, 7, 6.5) it took 64, 61 and 54 days to overcome the lack of resilience of the system in the interacting scenarios, compared to only 9, 8 and 7 days in the baseline scenarios. However, in the M=6 earthquake scenario, overcoming the lack of resilience in the interacting and baseline cases took only 13 and 4 days, respectively, principally because that the initial EPSS functionality loss at this earthquake intensity was much smaller than in the stronger earthquake scenarios, while the damage to the community built environment (i.e. drop in the electric power demand) was still significant. Therefore, the EPSS can cover the diminished power demand much sooner, even though it still took more than 50 days for the EPSS to fully restore its power generation capacity (Fig. 4). This illustrates the complex dynamics of electric power supply and demand during the post-earthquake recovery process. Another important observation is the long period of almost-constant *PPwoP* in the three larger magnitude interacting scenarios, particularly the 38-day interval (between the 7th and the 45th day) during the 7.5 interacting scenario during which the *PPwoP* remained virtually constant at 20%. This outcome can be very demanding for the population of the community. Finally, as shown by the quantile curves in Fig. 7(b), the uncertainty in the *PPwoP* data is quite large, indicating a need to identify and quantify the sources of this uncertainty and, possibly, devise recovery strategies aimed at reducing not only the median lack of resilience but also at reducing the uncertainty in achieving a satisfactory post-earthquake recovery of the EPSS-TS-Community system.

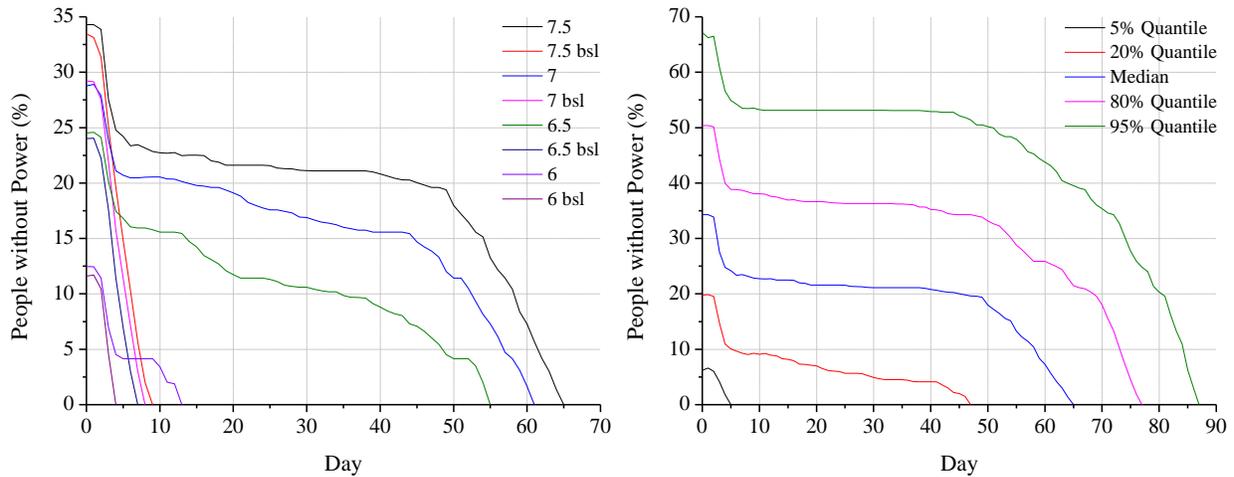

Fig. 7. Evolution of the *PPwoP* system resilience measure in case study scenarios without the **Administrator** agent (Fig. 7(a) comparison of median *PPwoP* graphs in different scenarios, Fig. 7(b) *PPwoP* data quantiles in the interacting M=7.5 earthquake scenario).

### *Case with the Administrator agent*

The effects of the interaction between the Community and the EPSS and TS are investigated by comparing the outcomes of the EPSS-TS-Community post-earthquake recovery simulations conducted with and without the **Administrator** agent. As stated in above, the community recovery performance thresholds were set to *PPwoP(RTC)* values of 10% (most demanding), 20% and 30% (least demanding) at the resilience check time of 3 days after the earthquake in the simulations with the **Administrator** agent.

The graphs showing the recovery of the median EPSS generation capacity *G(t)* in the interacting M=7.5 scenarios without and with the **Administrator** agent are shown in Fig. 8. The intervention of the community to enforce its recover priorities is quite effective for all *PPwoP* thresholds, particularly when the recovery performance target is the most demanding (*PPwoP*=10%) when it took 50 instead of 79 (Fig. 4) days to completely restore the 900 MW EPSS generation capacity.

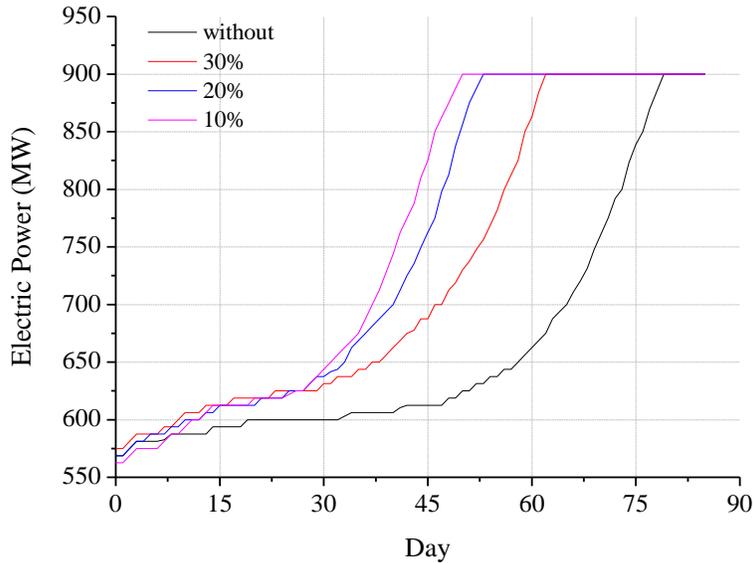

Fig. 8. Median generation capacity (*G*) recovery curves in the interacting scenarios case study simulations without and with the **Administrator** agent (with three *PPwoP* recovery performance thresholds).

Similar to Fig. 5, Fig. 9 presents the probability density histograms of the EPSS generation capacity data from the 2000 MC simulation instances of the interacting M=7.5 earthquake scenarios and the fitted normal distribution on the first and the 60th day after the earthquake. A comparison of Fig. 9(a) and Fig. 5(a) shows no differences because the **Administrator** agent is inactive as the resilience check time is set to the 3rd day after the earthquake. The data in Fig. 9(b) shows that the EPSS generation capacity was fully restored in less than 60 days in an overwhelming majority of the MC simulation instances. A comparison to the data in Fig. 5(b) shows that an **Administrator** agent intervention to enforce community recovery priorities can be very effective. A normal distribution to the entire dataset in Fig. 9(b) is clearly poor, even though it indicates a significant shift of the mean and a reduction of the standard deviation of the data.

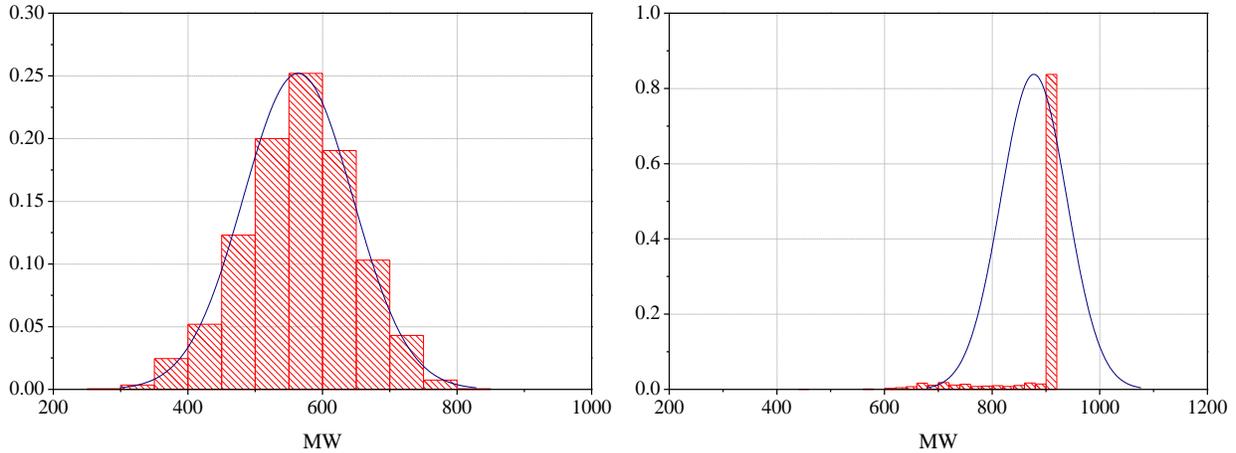

Fig. 9. Histogram of EPSS generation capacity (*G*) in the interacting M=7.5 case study scenario with the **Administrator** agent whose *PPwoP* recovery performance threshold is 10% (Fig. 9(a) First day, Fig. 9(b) 60[th] day).

The effect of **Administrator** agent intervention on the recovery of the TS is similar. As shown in the graphs of the median percentage of fully functioning bridges in the interacting M=7.5 earthquake scenarios in Fig. 10, the full recovery of TS is shortened by 26 days for the *PPwoP(RTC)* threshold value of 10%.

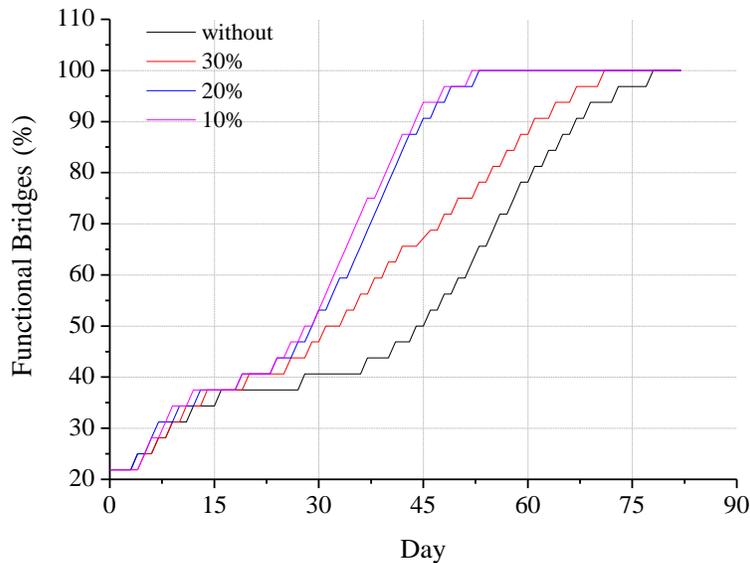

Fig. 10. Median of the fraction of functioning bridges in the TS in four interacting earthquake case study scenarios without and with the **Administrator** agent.

The recovery trajectory is roughly the same when this threshold is set to be 20%. However, as the **Administrator** agent becomes more tolerable (*PPwoP(RTC)* threshold is 30%), the effect of its intervention becomes less pronounced. It indicates that the community recovery performance threshold can have a significant impact on the recovery path of the EPSS-TS-Community system. The recovery performance objective should be set low enough so that the community can intervene and speed-up the recovery process, but not too low such that the recovery priorities of the CISs are completely neglected.

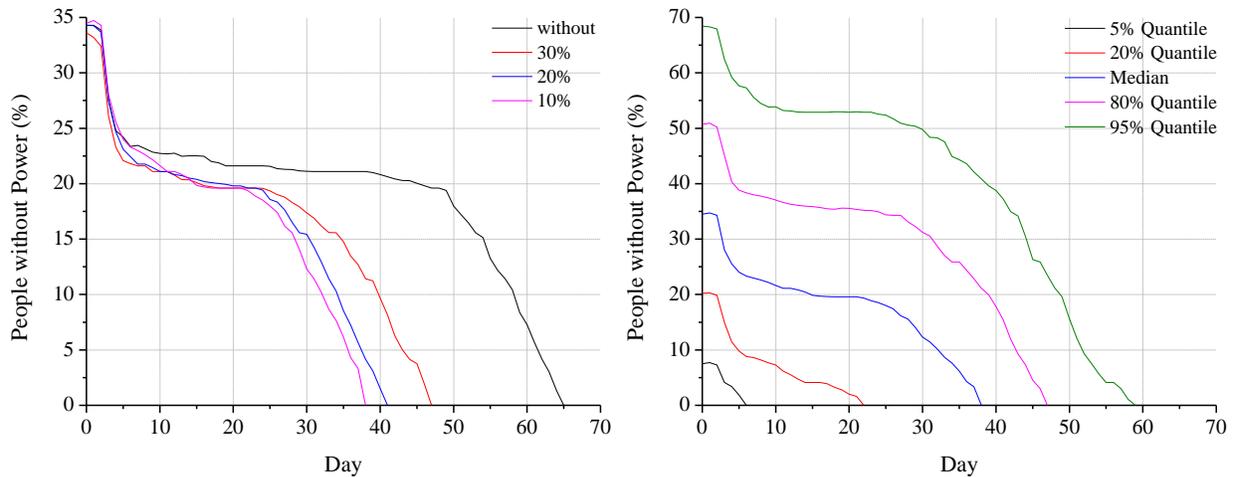

Fig. 11. Evolution of the *PPwoP* system resilience measure in interacting M=7.5 earthquake case study scenarios without and with the **Administrator** agent (Fig. 11(a) comparison of median *PPwoP* graphs for different *PPwoP* threshold values, Fig. 11(b) *PPwoP* data quantiles in the interacting M=7.5 scenario with the *PPwoP* threshold set to 10%).

The data on the rate of median *PPwoP* reduction shown in Fig. 11(a) also indicates that the intervention of the **Administrator** agent to speed up the recovery process can be very effective. The median *PPwoP* graphs in the simulations with the **Administrator** agent are significantly different compared to the case without the **Administrator** agent. Most importantly, the long horizontal "plateau" shown in Fig. 7 does not appear anymore, revealing that the community intervention achieved the goal of speeding up the recovery process. For example, in simulations with the *PPwoP(RTC)* performance threshold of 10% it only takes 38 days to alleviate the lack of resilience of the EPSS-TS-Community system, compared to 64 days it takes without community intervention, a 40% improvement. The quantile *PPwoP* curves shown in Fig 11(b) (analogous to Fig. 7(b)) indicate that the uncertainty of the simulation data remains large and is not affected by the activation of the **Administrator** agent. While the benefits of community intervention are clear, they do not come without costs. In this case study, the CIS operator costs increase directly, primarily through the doubling of the repair crew efficiency, but may also increase indirectly thorough the business losses incurred by prioritizing power dispatch to the more demanding by possibly less profitable segments of the community.

## Conclusions

The modern Civil Infrastructure System -Community System (CICS) is a dynamic and integrated socio-technical network. When such complex systems are stressed, as is the case in natural disasters like earthquakes, they become even more dynamic as interactions among the various elements of the system intensify as they cope with the consequences of the disaster and strive to restore themselves and the integrated system they are part of. A framework to evaluate the seismic resilience of modern CICSs was proposed in this paper. This framework is based on the compositional supply/demand approach to evaluate system resilience, conventional seismic vulnerability functions to evaluate component damage, conventional recovery functions to evaluate the recovery of the community built environment, and a novel agent-based model to simulate the recovery of multiple interdependent CISs after an earthquake. The principal features of the proposed agent-based model are the ability to model the individual participants in the CIS recovery process including their different strategies and capabilities, and the ability to model the interdependencies and interactions among these participants during the CICS recovery process.

The proposed agent-based seismic resilience evolution framework was exemplified in a case study on a virtual system comprising an electric power supply system, a transportation system and a community these two systems serve. The agent-based recovery model of the infrastructure systems features three agents, the **EPSS Operator, TS Operator,** and community **Administrator,** defined by a set of random variable attributes, parameters, and pre-planned recovery protocols. The agents interact in two ways: first, through the use of the transportation system by the EPSS repair crews, and second, through the intervention of the community to enforce its recovery performance objectives by speeding up and re-prioritizing the infrastructure system repairs. The operation of the model was verified and its behavior was examined in a series of Monte Carlo simulations involving a variety of earthquake magnitude and agent interaction scenarios.

The case study demonstrated that the agent-based modeling approach is quite suitable for modeling the recovery process of integrated civil infrastructure system and capable of representing the dynamic interactions between the participants of the recovery process. Extending the compositional supply/demand seismic resilience quantification framework by adding the agent-based recovery process models was also shown to be feasible and quite effective. The most remarkable outcome of the case study is the demonstration that a timely and well-planned intervention in the recovery process can be very effective in alleviating the post-earthquake lack of resilience resulting from the insufficient supply of civil infrastructure services to cover the community demands. The proposed agent-based seismic resilience evaluation model offers a way to investigate and design such interventions to guide the post-earthquake recovery process and increase the seismic resilience of modern CICSs.

The proposed agent-based recovery model is, however, not validated. While sufficiently granular data on the vulnerability and recovery of communities (Kang et al. 2018), infrastructure systems (Chang 2001, Nojima 2012) and combined community and infrastructure systems (Didier et al. 2017, Didier et al. 2018b) is becoming available, it still does not contain information about the infrastructure system operator and repair crew behavior to make it possible to realistically design and then validate the agent-based models. Until such data is collected, agent-based modeling of the recovery process can still be pursued in terms of developing multi-agent models (e.g. involving other infrastructure systems, private construction companies, different community and government entities), increasing the complexity of agent behavior (e.g. multiple repair crews, adaptable repair scheduling and re-scheduling), broadening the agent-based model to simulate the recovery of the community built environment in parallel with the recovery of the CISs, and introducing multiple recovery performance objectives to address not only the lack of resilience with respect to a certain civil infrastructure service, but also higher-level community functions. Such models could come closer to addressing the community resilience objectives outlined in (SPUR, 2009) and serve as a tool to evaluate benefits and costs of various strategies to increase the seismic resilience of communities.